\documentclass[%
 aps,
 prapplied,
 superscriptaddress,
 amsmath,amssymb,
 twocolumn,
 reprint,
 longbibliography
]{revtex4-2}

\usepackage{graphicx}
\usepackage{dcolumn}
\usepackage{bm}
\usepackage{physics}
\usepackage{siunitx}
\usepackage{xcolor}
\usepackage{hyperref}
\hypersetup{%
 colorlinks = true,%
 linkcolor = blue,
 citecolor = blue,
 urlcolor = blue
}

\newcommand{\II}{I\hspace{-0.05em}I}

\newcommand{\U}{\hat{U}}

\newcommand{\Ham}{\hat{\mathcal{H}}}
\begin{document}


\title{Floquet engineering using pulse driving in a diamond two-level system \\ under
large-amplitude modulation}

\author{Shunsuke Nishimura}
\affiliation{Department of Physics, The University of Tokyo, Bunkyo-ku, Tokyo 113-0033, Japan}
\author{Kohei M. Itoh}%
\affiliation{School of Fundamental Science and Technology, Keio University, Kohoku-ku, Yokohama 223-8522, Japan}
\author{Junko Ishi-Hayase}%
\affiliation{School of Fundamental Science and Technology, Keio University, Kohoku-ku, Yokohama 223-8522, Japan}
\author{Kento Sasaki}%
\affiliation{Department of Physics, The University of Tokyo, Bunkyo-ku, Tokyo 113-0033, Japan}
\affiliation{School of Fundamental Science and Technology, Keio University, Kohoku-ku, Yokohama 223-8522, Japan}
\author{Kensuke Kobayashi}%
\affiliation{Department of Physics, The University of Tokyo, Bunkyo-ku, Tokyo 113-0033, Japan}
\affiliation{Institute for Physics of Intelligence, The University of Tokyo, Bunkyo-ku, Tokyo 113-0033, Japan}
\affiliation{Trans-scale Quantum Science Institute, The University of Tokyo, Bunkyo-ku, Tokyo 113-0033, Japan}

\date{\today}

\begin{abstract}
The nitrogen-vacancy (NV) center in a diamond is a promising platform for Floquet engineering. Using the synchronized readout, we investigate the NV center's Floquet state driven by the Carr-Purcell sequence in a large-amplitude AC magnetic field. We observe the dynamics represented as Bessel functions up to 211th orders high in a systematic and quantitative agreement with the theoretical model. Furthermore, numerical calculations show that the effect of finite pulse duration and error limits the modulation amplitude available for Floquet engineering. This work provides an approach to precisely investigate Floquet engineering, showing the extendable range of modulation amplitude for two-level systems.
\end{abstract}
\maketitle
\section{Introduction}
Floquet theory gives a general formalism for the dynamics of quantum systems subject to periodic driving~\cite{shirley1965solution,sambe1973steady}. 
Recently, utilizations of various periodically-driven quantum dynamics have been attracting keen attention as Floquet engineering~\cite{OkaPRB2009,malz2021topological,martin2017topological,eckardt2017colloquium,weitenberg2021tailoring,oka2019floquet,lukin2020spectrally}.
A thorough investigation of the states produced by diverse periodic drivings is essential for further development.
In this context, the Floquet engineering using pulse driving has aroused interest as a stage for Hamiltonian engineering~\cite{haeberlen1968coherent,choi2020robust} as well as from perspectives such as discrete-time crystals physics~\cite{choi2017observation,randall2021many, hengyun2020quantum, zhang2017observation,beatrez2021floquet} and quantum sensing~\cite{de2011single,lang2015dynamical,meinel2021heterodyne}.

One of the fundamental interests in Floquet engineering is the fidelity of an experimental system to the physical model under a large-amplitude periodic modulation~\cite{ashhab2007two}. 
The higher-order oscillatory response often appears nonlinearly in this regime, hindering quantitative interpretation.   
Many researchers reported the peculiar dynamics in such a large-amplitude regime, for example, Rabi oscillations~\cite{fuchs2009gigahertz,scheuer2014precise,rao2017nonlinear,wang2021observation,deng2015observation,dai2017quantum}, multi-frequency absorption~\cite{saito2006parametric}, Landau-Zener interferometry~\cite{oliver2005mach,berns2006coherent,berns2008amplitude}, Carr-Purcell (CP) sequence~\cite{kotler2013nonlinear,zopes2017high}, and cavity couplings~\cite{niemczyk2010circuit,forn2010observation,yoshihara2017superconducting,yoshihara2017characteristic,yoshihara2018inversion}. However, it is still unknown to what extent higher-order nonlinear responses in such a regime can be accurately understood beyond qualitative observations.

In this paper, we focus on a two-level system in a diamond nitrogen-vacancy (NV) center driven by the CP sequence to enter the Floquet regime.
We use AC magnetometry, where an AC magnetic field is applied to a single NV center~\cite{MSH+08,HHC+10,dRD+11,NDH+11} driven by the CP sequence, as a test platform of Floquet engineering.
Adopting the synchronized readout technique~\cite{schmitt2017submillihertz, boss2017quantum,glenn2018high-resolution}, we investigate the nonlinear response caused by the phase accumulation in a large-amplitude modulation of the AC magnetic field and, thus, prove that the response obeys the Bessel functions up to as high as 211th orders, in accurate agreement with the theory.
This work provides a solid foundation for developing pulse-driven Floquet engineering.

This paper is organized as follows:
We first define the Floquet picture of pulse drives in Section~\ref{sec:floquet}. 
In Section~\ref{sec:syncronized}, we then describe the principle of extracting the Floquet state by the synchronized readout.
After explaining the measurement setup in Section~\ref{sec:experiment}, we show the experimental results in Section~\ref{sec:results}.
In Section~\ref{sec:discussion}, we discuss the implication and potential applications of the present achievement and conclude in Section~\ref{sec:conclusion}.

\section{Floquet Picture}\label{sec:floquet}
\begin{figure}[!b]
\includegraphics[width=0.95\hsize]{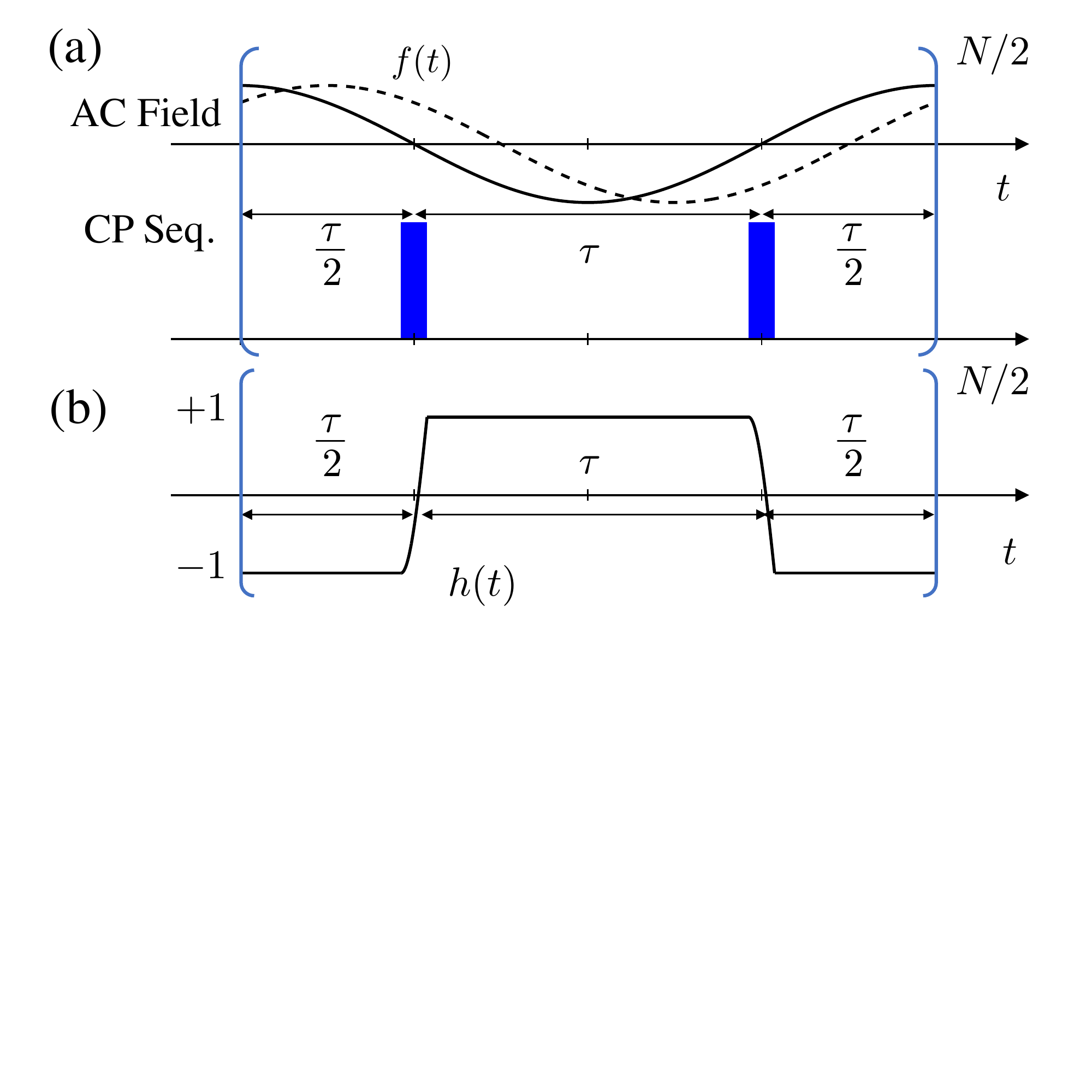}
\caption{(a) Diagram of the AC field and the pulse cycle of the CP sequence (CP Seq.). $N$ denotes the total number of pulses. In this case, the number of repetitions equals $N/2$. The dashed curve represents the AC field with different phase. Our method can also apply to AC field of arbitrary phase as discussed in Sec.~\ref{subsec:Precession_of_the_Floquet}. 
(b) Corresponding modulation function $h(t)$ over time. See Eq.~\eqref{ht}.}
\label{fig:cp2}
\end{figure}

We start with the Floquet picture of the pulse-driven systems in the present study. 
To realize pulse-driven Floquet engineering, strict periodicity is necessary.
For example, the period of the pulse sequence and the other modulations, such as AC field drives, must coincide precisely to realize a periodic Hamiltonian. 
As a representative of such a pulse-driven Floquet system, we consider the CP sequence with AC field modulation where the interpulse delay and AC field half-period match, as depicted in Fig. \ref{fig:cp2}(a).

We divide the driving force into two parts, one by the CP sequence and the other by the AC field modulation, as shown in Fig.~\ref{fig:cp2}(a).
After defining the Floquet state and the Floquet mode in Sec.~\ref{subsec:Floquet_picture}, we extract the time-dependent eigenstates of the CP sequence as a basis in the absence of the AC field in Sec.~\ref{subsec:Floquet_state_driven_by_CP_sequence}. 
Then, in Sec.~\ref{subsec:Precession_of_the_Floquet}, we take the convolution of the time evolution of these eigenstates and the AC field modulation to calculate the total unitary operator in the presence of the AC field drive.

\subsection{Definition}\label{subsec:Floquet_picture}
When a Hamiltonian $\hat{H}(t)$ has a periodicity such that $\hat{H}(t)=\hat{H}(t+T)$, the solution $\ket{\Psi_{F}(t)}$ of the time-dependent Schr\"odinger equation ($\hbar=1$),
\begin{align}
i\frac{\partial}{\partial t} \ket{\Psi_\text{F}(t)} = \hat{H}(t)\ket{\Psi_\text{F}(t)},
\label{eq:schr}
\end{align}
can be described by the Floquet theory. Accordingly, $\ket{\Psi_{F}(t)}$ can be expressed with periodic states $\ket{\Phi_{\alpha}(t+T)} = \ket{\Phi_{\alpha}(t)}$
in the following form:
\begin{align}
\ket{\Psi_\text{F}(t)} = \sum_{\alpha} c_{\alpha} e^{i \epsilon_{\alpha}t } \ket{\Phi_{\alpha}(t)} ,\quad c_{\alpha}\in \mathbb{C}
\label{eq:fstate}
\end{align}
Here, we call $\ket{\Psi_\text{F}(t)}$ Floquet state, $\ket{\Phi_{\alpha}(t)} $ Floquet mode, and $\epsilon_{\alpha}$ quasienergy. 
The Floquet mode and quasienergy are determined by the eigenvalue problem of a time-independent Floquet Hamiltonian $\Ham_\alpha$,
\begin{align}
\Ham_{\alpha} &= \hat{H}(t) - i\frac{\partial}{\partial t},\\
\Ham_{\alpha} \ket{\Phi_{\alpha}(t)} &= \epsilon_{\alpha} \ket{\Phi_{\alpha}(t)}.
\label{eq:fmode}
\end{align}
The $T$-periodicity of the Floquet mode straightforwardly explains the stroboscopic response of the Floquet state. Such formalism helps an easy understanding of a periodically driven system.

\subsection{Floquet state driven by CP sequence}\label{subsec:Floquet_state_driven_by_CP_sequence}

We focus on the $m_S=0$ and $-1$ states of the NV center in a static magnetic field, which is a two-level system treated in the present experiment~\cite{MSH+08,HHC+10,dRD+11,NDH+11}. 
We express them as $\ket{0}$ and $\ket{1}$, respectively. 
The spin projection operator $\hat{S}_z$ is then expressed as \begin{equation}
    \hat{S}_z = -\ket{1}\bra{1}.
\end{equation}
We derive the Floquet state for the CP sequence in the absence of the AC field in this subsection and then show the effect of the AC field in the following subsection.
We assume that the microwave pulse is rectangular and resonant to the NV center.
The Hamiltonian for the CP sequence in the rotational coordinate system under the rotating wave approximation is given as,
\begin{align}
\hat{H}^{\mathrm{CP}} (t)  
=& 
\begin{cases}
\Omega_\text{R} \hat{S}(\phi_1) & \frac{\tau-t_{\pi}}{2}+2n\tau \leq t < \frac{\tau+t_{\pi}}{2}+2n\tau \\
\Omega_\text{R} \hat{S}(\phi_2) & \frac{3\tau-t_{\pi}}{2}+2n\tau \leq t < \frac{3\tau+t_{\pi}}{2}+2n\tau \\
0 & \mathrm{otherwise},
\end{cases}\notag\\
\quad \notag\\
&\text{for}~n = 0,1,\cdots, (N/2-1)
\label{eq:hcp}
\end{align}
where $\Omega_R=\pi/t_{\pi}$ is the Rabi frequency, $\hat{S}(\phi_i)=\frac{1}{2}(e^{i\phi_i}\ket{1}\bra{0}+e^{-i\phi_i}\ket{0}\bra{1})$, $\phi_i$ is the pulse phase ($i=1,2$), $t_{\pi}$ is the $\pi$ pulse duration, $\tau$ is the interpulse delay and $N$ is the number of $\pi$ pulses.

The Hamiltonian [Eq.~\eqref{eq:hcp}] is modified according to the pulse phase series used in the CP sequence. 
For a technical reason, we first consider the situation that the pulse duration takes the finite span $t_\pi$. 
Since it has a periodicity of $T=2\tau$, a quantum state under the Hamiltonian is a Floquet state.
When the initial state is $\ket{\psi_0(t=0)}=\ket{0}$, the time evolution in one cycle $t=[0,2\tau]$ of the CP sequence is obtained as,
\small
\begin{align}
\ket{\psi_0(t)}
&=
\begin{cases}
\ket{0} & 0 \leq t < \frac{\tau-t_{\pi}}{2} \\
\U_p(\phi_1,t,\frac{\tau-t_{\pi}}{2}) \ket{0} & \frac{\tau-t_{\pi}}{2} \leq t < \frac{\tau+t_{\pi}}{2} \\
-i e^{-i\phi_1} \ket{1} & \frac{\tau+t_{\pi}}{2} \leq t < \frac{3\tau-t_{\pi}}{2} \\
-i e^{-i\phi_1} \U_p(\phi_2,t,\frac{3\tau-t_{\pi}}{2}) \ket{1} & \frac{3\tau-t_{\pi}}{2} \leq t < \frac{3\tau+t_{\pi}}{2} \\
-e^{-i(\phi_1-\phi_2)} \ket{0} & \frac{3\tau+t_{\pi}}{2} \leq t < 2\tau,
\end{cases}
\label{eq:psi0}
\end{align}
\normalsize
where $\U_p(\phi,t_f,t_i)$ represents the pulse operation during $t \in [t_i,t_f]$ with phase $\phi$, which is given as
\small
\begin{align}
\U_p(\phi,t_f,t_i) 
&=\exp[-i \frac{\pi}{t_{\pi}} (t_f-t_i) \hat{S}(\phi)]\notag\\
&= \cos\qty[\frac{\pi}{2t_{\pi}}(t_f-t_i)]\bm{1} - 2i\sin\qty[\frac{\pi}{2t_{\pi}}(t_f-t_i)]\hat{S}(\phi).
\label{up}
\end{align}
\normalsize
The state $\ket{\psi_0(t)}$ is an eigenstate of the Hamiltonian [Eq.~\eqref{eq:hcp}], which satisfies Eq.~\eqref{eq:schr}, and therefore a Floquet state. Similarly, the state 
\begin{equation}
 \ket{\psi_1(t)}=(\ket{1}\bra{0}+\ket{0}\bra{1})\ket{\psi_0(t)},
 \label{eq:psi1}
\end{equation} 
which is orthogonal to $\ket{\psi_0(t)}$ is also a Floquet state. These two states are, in nature, normalized and orthogonal to each other. 
Furthermore, the basis of these two states $\ket{\psi_0 (t)}$ and $\ket{\psi_1 (t)}$ span the whole Floquet state represented in Eq.~\eqref{eq:fstate}.

The global phase offset $\phi_1-\phi_2+\pi$ is accumulated to the quantum state every cycle of the CP sequence as in Eq.~\eqref{eq:psi0}. 
This phase does not change the observables in this two-level system.
Thus, for simplicity, we fix the pulse phase condition to be $\phi_1-\phi_2+\pi = 0$.
We denote  $\ket{\Phi_0^{\mathrm{CP}} (t) }$ and $\ket{\Phi_1^{\mathrm{CP}} (t) }$
as the special case of $\ket{\psi_0(t)}$ and $\ket{\psi_1(t)}$, respectively, for that condition.
Then, they satisfy the following periodic conditions,
\begin{align}
 \ket{\Phi_0^{\mathrm{CP}}(t+2\tau) } &= \ket{\Phi_0^{\mathrm{CP}} (t) },\\
 \ket{\Phi_1^{\mathrm{CP}}(t+2\tau) } &= \ket{\Phi_1^{\mathrm{CP}} (t) }.
\end{align}
These states are Floquet modes defined in Eq.~\eqref{eq:fstate} with quasienergy $\epsilon_\alpha^{\mathrm{CP}} =0$.
With these Floquet modes, the Floquet state $\ket{\Psi_\text{F}(t)}$ is expressed by using appropriate coefficients $c_{0}'$ and $c_{1}'$ such that
\begin{equation}
 \ket{\Psi_\text{F}(t)} = 
 c_{0}' \ket{\Phi_0^{\mathrm{CP}} (t) }
 +
 c_{1}' \ket{\Phi_1^{\mathrm{CP}} (t) }.
 \label{eq:cpbasis}
\end{equation}
Now, without losing the essence of the Floquet picture, we can set the $\pi$ pulse duration $t_\pi \to 0$. 
The CP pulse Hamiltonian Eq.~\eqref{eq:hcp} within one period~$[0,2\tau]$ can be rewritten using the Dirac delta $\delta(t)$ as
\begin{equation}
\hat{H}^{\mathrm{CP}} (t) = \pi\delta(t-\tau/2) \hat{S}(\phi_1) + \pi\delta(t-3\tau/2) \hat{S}(\phi_2).
\label{eq:hcp_pulse}
\end{equation}
Accordingly, the Floquet modes of the CP sequence are represented as the eigenstates of $\hat{H}^{\mathrm{CP}} (t)$ such that
\begin{align}
\ket{\Phi_0^{\mathrm{CP}} (t) } &= \frac{1-h(t)}{2}\ket{0} + \frac{1+h(t)}{2}\ket{1}, \label{eq:Phi0_pulse}\\
\ket{\Phi_1^{\mathrm{CP}} (t) } &= \frac{1+h(t)}{2}\ket{0} + \frac{1-h(t)}{2}\ket{1}, \label{eq:Phi1_pulse}
\end{align}
where $h(t)$ is the modulation function
\begin{align}
h(t) &= 
\begin{cases}
-1 & 0 < t < \frac{\tau}{2} , \quad \frac{3\tau}{2} < t < 2\tau \\
1 & \frac{\tau}{2} < t < \frac{3\tau}{2}.
\end{cases}
\label{ht}
\end{align}
which is depicted in Fig.~\ref{fig:cp2}(b).

Eqs.~\eqref{eq:psi0} and \eqref{eq:psi1} tell that since the pulse phases, $\phi_1$ and $\phi_2$, do not affect the overall dynamics of the Floquet state other than the global phase, we can arbitrarily choose them. Thus, in our experiment, we employ a pulse cycle containing multiple phases for robust operation as detailed in Sec.~\ref{subsec:protocol}.

\subsection{Precession of the Floquet state driven by CP sequence in the AC magnetic field}
\label{subsec:Precession_of_the_Floquet}
We consider the case where an arbitrary AC magnetic field $f(t)$ is applied in the direction of the symmetry axis of the NV center.
Generally, a periodic function $f(t)$ with a period $T=2\tau$ can be expanded in the Fourier series as,
\begin{equation}
f(t) = a_0 + \sum_{k = 1}^\infty \qty[ a_k \cos(k\omega t) + b_k \sin(k\omega t) ],
\label{eq:ft}
\end{equation}
where $\omega=\pi/\tau$. A Hamiltonian under the AC magnetic field is given in the Hilbert space that the spin operator basis spans:
\begin{equation}
 \hat{H}^{\mathrm{AC}} (t) = f(t)\hat{S}_z = -f(t)\ket{1}\bra{1}. \label{eq:hac}
\end{equation}
The total Hamiltonian $\hat{H}$ for the quantum state driven by the CP sequence in the AC magnetic field is given by,
\begin{equation}
 \hat{H}(t) = \hat{H}^{\mathrm{AC}} (t) + \hat{H}^{\mathrm{CP}} (t),
 \label{eq:htot}
\end{equation} 
which is represented on the same basis. 

From here, in order to clarify the effect of the AC magnetic field to the Floquet state, 
we write down the above Hamiltonian $\hat{H}$ on the basis of 
the Floquet mode $\ket{\Phi_i(t)^{(\text{CP})}},\, i=0,1$ in Eqs.~\eqref{eq:Phi0_pulse} and \eqref{eq:Phi1_pulse}. 
We use the following unitary transformation $\hat{U}_t(t)$ to represent the frame change $\ket{i} \to \ket{\Phi_i(t)^\text{CP}}$:
\begin{equation}
\hat{U}_t(t) = \ket{\Phi_0^\text{CP}(t)}\bra{0} + \ket{\Phi_1^\text{CP}(t)}\bra{1}. 
\end{equation}
This unitary transformation represents the driving of the spin state by $\hat{H}^{\mathrm{CP}} (t)$. Since $\ket{\Phi_i ^{\mathrm{CP}} (t) }$ are orthonormal eigenstates of $\hat{H}^{\mathrm{CP}} (t)$ [Eq.~\eqref{eq:hcp_pulse}] that satisfy the time-dependent Shr\"odinger equation [Eq.~\eqref{eq:schr}], the identity $\displaystyle{i{\pdv{\hat{U}_t}{t}} (t)  = \hat{H}^\text{CP}(t) \hat{U}_t (t)}$ holds.
Then we find, 
\begin{equation}
i\U_t^\dagger(t) \frac{\partial \U_t}{\partial t} (t)  = \hat{H}^{\mathrm{CP}} (t).
\end{equation}
Thus, in this frame, the total Hamiltonian $\breve{H}(t)$ is given as
\begin{align}
\breve{H}(t) &=  \hat{U}_t^\dagger(t) \hat{H} (t) \hat{U}_t (t)- i\U_t^\dagger(t) \frac{\partial \U_t}{\partial t} (t)\notag\\
&= -f(t)
\frac{1+h(t)}{2} \ket{0} \bra{0} - f(t)\frac{1-h(t)}{2} \ket{1} \bra{1} .
\label{eq:htot_cp}
\end{align}
In the above, we utilize the following identities: 
\[
 \qty(\frac{1\pm h}{2})^2 = \frac{1\pm h}{2},\quad \qty(\frac{1 + h}{2})\qty(\frac{1 - h}{2}) = 0.
\]

In the presence of the AC field, the precession frequencies of the two states $\ket{0}, \ket{1}$ are given as,
\begin{equation}
\xi_{0}^{\mathrm{CP}} (t) = -f(t)\frac{1 + h(t)}{2}, \quad \xi_{1}^{\mathrm{CP}} (t) = -f(t)\frac{1 - h(t)}{2},
\label{eq:xi}
\end{equation}
 respectively. 
 Thus, the phase accumulations during $t=[0,2\tau]$ that correspond to the quasienergies are calculated as
 \begin{align}
     \phi_0 &= \int_0^{2\tau} \xi_0^{\mathrm{CP}}(t)dt 
     = -a_0\tau - \sum_{j = 0}^\infty a_{2j+1} \frac{2(-1)^{j}}{(2j +1)\omega}, \\
     \phi_1 &= \int_0^{2\tau} \xi_1^{\mathrm{CP}}(t)dt 
     = -a_0\tau + \sum_{j = 0}^\infty a_{2j+1} \frac{2(-1)^{j}}{(2j +1)\omega}.
 \label{phiu}
 \end{align}
Here, we use the symmetry $h(t)=h(2\tau-t)$ and the property of convolution $(h*f)(2\tau)=\int_0^{2\tau} h(2\tau-t)f(t)dt$ and the Fourier series of $h(t)$ given as
\begin{align}
h(t) = \sum_{j=0}^{\infty} \frac{4}{\pi} \frac{(-1)^{j}}{(2j +1)} \cos[ (2j+1)\omega t ].
\label{eq:htft}
\end{align}
Since $h(t)$ consists only of the odd-order cosine part of the Fourier composition, 
only the matched components of $f(t)$ [$a_{2j+1}$ in Eq.~\eqref{eq:ft}] contribute to the dynamics, as derived from the property of the convolution.
The dynamics $\U$ in the overall sequence $t=[0,2\tau \times (N/2)]$ is obtained as,
\begin{align}
\U &= 
e^{i\phi_0\qty(N/2)} \ket{0 } \bra{0} +e^{i\phi_1\qty(N/2)} \ket{1} \bra{1},
\label{eq:U}
\end{align}
and here,
\begin{equation}
 \phi_\text{acq} = \frac{N}{2}(\phi_1 - \phi_0) =  \sum_{j = 0}^\infty a_{2j+1} \frac{2N(-1)^{j}}{(2j +1)\omega}, 
 \label{eq:phiacq}
\end{equation}
is the phase difference between Floquet modes.
In the stroboscopic measurement at the end of the CP sequence, the state $\ket{0}(\ket{1})$ is experimentally indistinguishable from the state $\ket{\Phi_0^{\mathrm{CP}} (t) }(\ket{\Phi_1^{\mathrm{CP}} (t) })$. 
Thus, the treatment that exploits the frame change $\ket{i} \to \ket{\Phi_{i}^{\mathrm{CP}} (t) }$ can be used to describe the Floquet state of the pulsed drives.

The Floquet state dynamics is explained by the convolution of the modulation function $h(t)$ and the target AC field $f(t)$.
The CP sequence can be understood as engineering to extract only the cosine of the odd harmonic components of $f(t)$ as in Eq.~\eqref{phiu}.
By designing the modulation function $h(t)$ by adjusting $\tau$ appropriately, we can extract any frequency components of the AC field.

\section{Synchronized readout}\label{sec:syncronized}
We have obtained the phase dynamics of the Floquet state driven by the CP sequence in the AC magnetic field. 
The dynamics is reflected in the phase acquisition $\phi_\text{acq}$.
We next formulate the principle of retrieving the Floquet state by the so-called synchronized readout, which we focus on in this study. 

\subsection{Phase dynamics measurement of a Floquet state by the synchronized readout}
\label{subsec:Principle_of_measuring}
\begin{figure}[tbp]
 \includegraphics[width=0.95\linewidth]{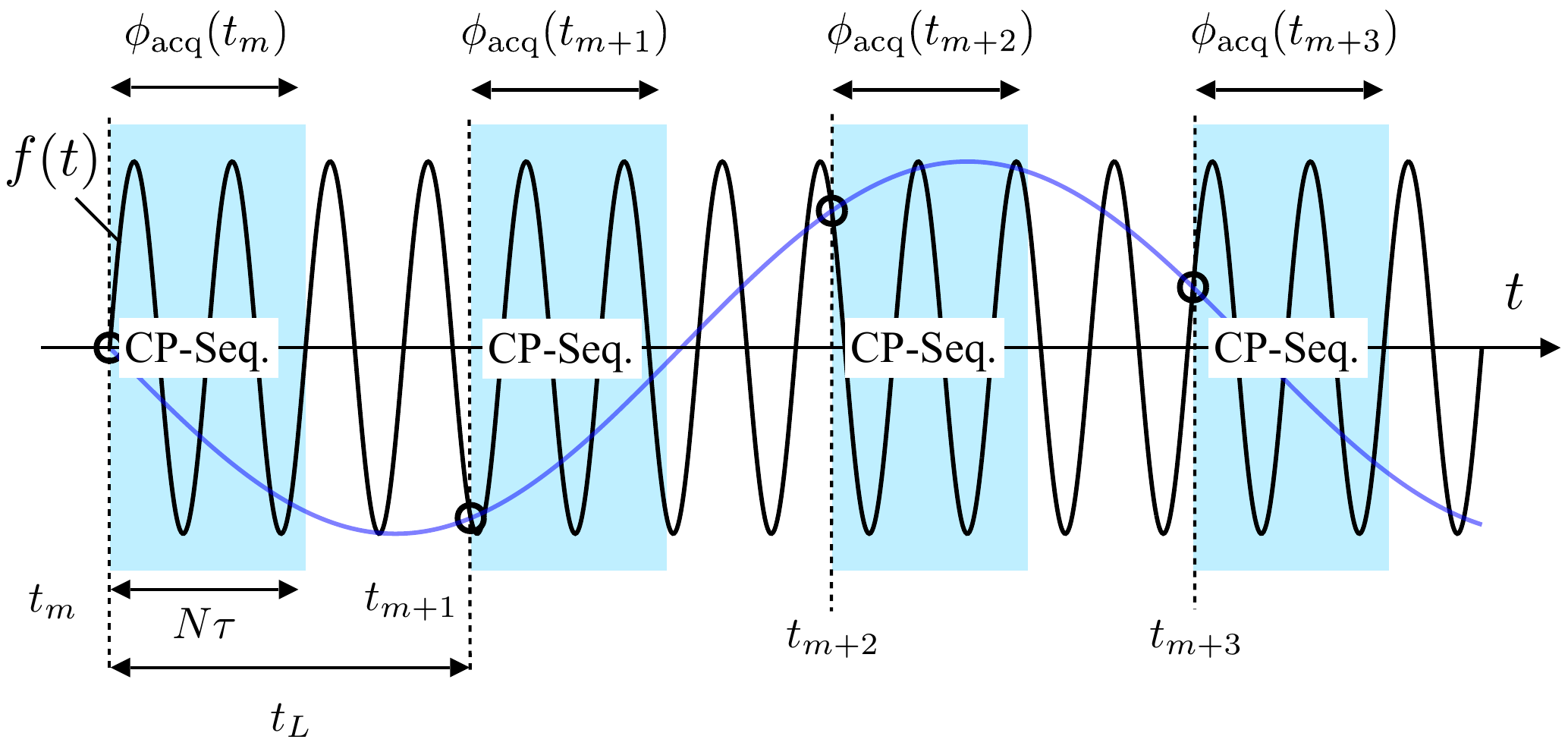}
 \caption{
Schematic of the synchronized readout. The blue-shaded blocks marked ``CP-Seq.'' represent the sequence shown in Fig.~\ref{fig:cp2}(a).
The blue line represents the systematic AC field phase shift in each CP sequence. 
The period of this blue-lined modulation is determined by both the period of the AC field ($2\tau$) and the time interval $t_L$ of the synchronized readout, and is obtained from the undersampling condition (see Appendix \ref{sec:undersample}).
 }\label{fig:sync_readout}
\end{figure}

We use the initial state 
\begin{align}
 \ket{\Psi_\text{F}(t=0)} =(\ket{0}-i\ket{1})/\sqrt{2},
\end{align} 
to interferometrically measure the difference of the phase acquisition in the Floquet modes [see Eq.~\eqref{eq:cpbasis}].
Setting the phase of the readout $\pi/2$ pulse as $\phi=\frac{\pi}{2}$, the probability of detecting the quantum state as $\ket{0}$ is obtained as
\begin{align}
P_0 &= \frac{1 + \sin \phi_\text{acq} }{2}.
\label{eq:p0}
\end{align}
The phase accumulation is experimentally accessible by probing $P_0$.

We systematically investigate $\phi_\text{acq}$ by using the synchronized readout technique. 
As shown in Fig.~\ref{fig:sync_readout}, this technique repeats the CP sequence at a time interval of $t_L$.
We redefine the time origin $t$ as the timing of the first CP sequence of the synchronized readout; the start time of the $m $th CP sequence ($m= 1,2,3,\cdots$) is $t_m = (m-1) t_L$.
The time shift in the CP sequences causes a systematic phase shift to the AC field [Eq.~\eqref{eq:ft}] such that
\small
\begin{align}
f(t+t_m) &= a_0 + \sum_{k = 1}^\infty \qty[a_k \cos(k\omega t_m) + b_k \sin(k\omega t_m)] \cos(k\omega t) \notag\\
&+ \qty[-a_k \sin(k\omega t_m) + b_k \cos(k\omega t_m)] \sin(k\omega t). 
\label{eq:ftm}
\end{align}
\normalsize
Then, $\phi_\text{acq}(t_m)$ in the $m$th CP sequence can be derived 
from Eq.~\eqref{eq:phiacq} and the Fourier expansion of $f(t)$~[Eq.~\eqref{eq:ft}] as follows:
\begin{align}
\phi_\text{acq}(t_m)
&= \sum_{j = 0}^\infty \frac{2N(-1)^{j}}{(2j +1)\omega} \{a_{2j+1} \cos[(2j+1)\omega t_m] \notag\\
&+ b_{2j+1} \sin[(2j+1)\omega t_m]\}.\label{eq:phiacq_sync}
\end{align}
The Fourier series for $t_m$ of $\phi_\text{acq}(t_m)$ contains all odd harmonic components of $f(t)$.
We can obtain each odd harmonic component of $f(t)$ as the spectrum of the time-series readout result $\qty{P_0(t_m)}$.

\subsection{Fourier decomposition to extract experimental signal}
\label{subsec:Method_to_compare}

Our experiment observes the dynamics in the simplest sinusoidal AC magnetic field.
It corresponds to the case where all the Fourier series components 
except for $k=1$ in Eqs.~(\ref{eq:ft}) and (\ref{eq:ftm}) are zero, that is, 
\small
\begin{align}
f(t) &= \gamma b_\text{ac} \cos(\omega_\text{ac} t + \phi_\text{ac}) \notag\\
&= \gamma b_\text{ac} \cos(\phi_\text{ac})\cos(\omega_\text{ac} t) - \gamma b_\text{ac} \sin(\phi_\text{ac})\sin(\omega_\text{ac} t),
\end{align}
\normalsize
where $b_\text{ac}$, $\omega_\text{ac}\,(=\pi/\tau)$, and $\phi_\text{ac}$ are the amplitude, the frequency, and the phase of the AC field, respectively, and $\gamma = 2\pi\times 28~\mathrm{rad~GHz/T}$ is the gyromagnetic ratio.
According to Eq.~\eqref{eq:phiacq_sync}, the phase acquisition at the $m$th CP sequence is obtained as
\small
\begin{align}
\phi_\text{acq}(t_m) 
&= \frac{2N \gamma b_\text{ac} }{\omega_\text{ac}} [\cos\,\phi_\text{ac}\cos(\omega_\text{ac} t_m) - \sin\,\phi_\text{ac} \sin(\omega_\text{ac} t_m)]\notag\\
&= \frac{2N \gamma b_\text{ac} }{\omega_\text{ac}} \cos(\omega_\text{ac} t_m + \phi_\text{ac}).
\end{align}
\normalsize
The signal at the $m$th readout is obtained as
\small
\begin{align}
P_0(t_m) &= \frac{1}{2} + \frac{1}{2}\sin\qty[\frac{2N \gamma b_\text{ac} }{\omega_\text{ac}} \cos(\omega_\text{ac} t_m + \phi_\text{ac})]\notag\\
&=\frac{1}{2}+\frac{1}{2}\mathrm{Im}
 \qty[\exp(i \frac{2N \gamma b_\text{ac} }{\omega_\text{ac}}
 \cos(\omega_\text{ac} t_m + \phi_\text{ac}))].
\end{align}
\normalsize
The Fourier series expansion of the signal for $t_m$ is obtained as,
\begin{align}
P_0(t_m) &= \frac{1}{2} + \sum_{k:\text{odd}} A_k \cos(k\omega_\text{ac} t_m + \phi_k), \\
A_k &= J_{k}\qty(2N \frac{\gamma b_\text{ac}}{\omega_\text{ac}}), \label{eq:Ak} \\
\phi_k &= k\qty(\frac{N\pi}{2} +\phi_\text{ac} -\frac{\pi}{2}) - \frac{\pi}{2},
\end{align}
where $J_k$ is a Bessel function of the first kind of $k$th order.
Here, we use the Jacobi--Anger identity
\begin{equation}
 e^{ ia\cos \varphi }=\sum _{n\in \mathbb{Z}} i^{n}J_{n}(a)e^{in\varphi}.
\end{equation}
The frequency resolution of the discrete Fourier transformation (DFT) spectrum can be set arbitrarily 
by the inverse of the total duration of the synchronized readout.
Thus, with high accuracy, we can investigate the response dependent on the amplitude and phase of the AC field.
According to Eq.~\eqref{eq:Ak}, higher-order Bessel functions take a non-zero value and oscillate when the magnetic field amplitude increases. 
For a sufficiently weak magnetic field, the behavior of the $A_k$ is linear to $b_{ac}$.
The threshold above which the response is no longer regarded linear should be
\begin{equation}
 b_{ac} \approx \frac{\pi}{2}\frac{\omega_{ac}}{2N\gamma}.
 \label{baclin}
\end{equation}

Our primary concern of the present work is to determine the degree to which the AC field amplitude is consistent with the DFT amplitude given by Eq.~\eqref{eq:Ak} as a crucial test of the robustness of this Floquet state against large amplitude modulation.

\section{Experiments}\label{sec:experiment}

This section describes our experimental condition.
We measure a single NV center in a \II a diamond substrate (of electrical grade from Element Six) using a homemade confocal system~\cite{misonou2020construction}.

\begin{figure}
\includegraphics[width=\hsize]{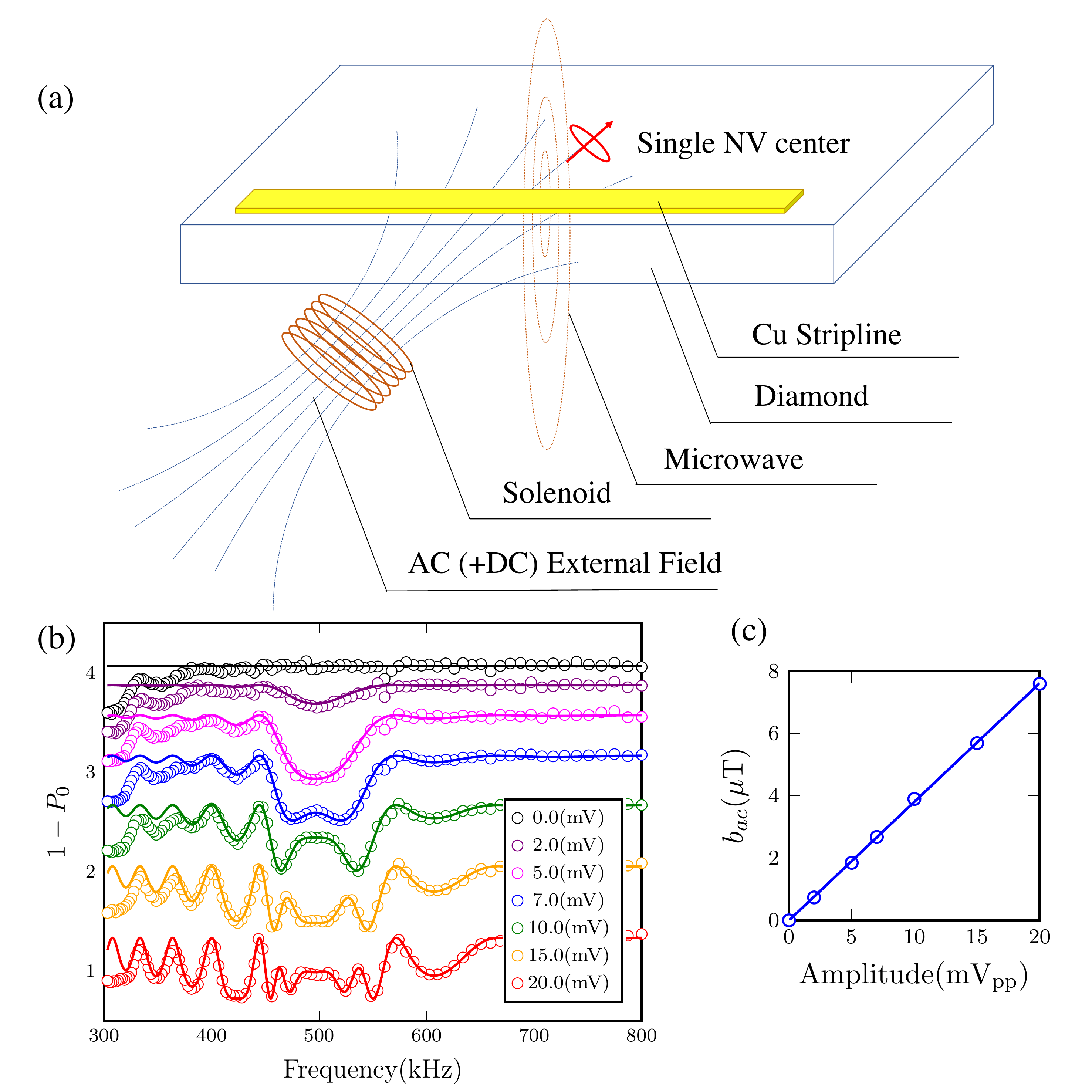}
\caption{
(a) Schematic of our experimental setup. 
(b) AC magnetometry using CP sequence. The target AC magnetic field frequency is 500.1~kHz.
For clarity, the data of each condition are shifted vertically.
(c) The relation between input voltage amplitude and AC field amplitude obtained from the fitting of (b).
}\label{fig:setup}
\end{figure}
\subsection{Setups}
Figure~\ref{fig:setup}(a) is the schematic of our experimental setup. We apply a static bias magnetic field of around 30~mT by a Neodymium magnet beneath the diamond [not shown in Fig.~\ref{fig:setup}(a)] to lift the spin degeneracy of the energy levels.
For the generation of a microwave field to control the NV center, we use a copper stripline of approximately $1~\si{\micro\meter}$-thick and $200~\si{\micro\meter}$-wide fabricated right on the surface of the diamond using an electron-beam deposition.
The microwave pulse waveform is generated by IQ modulation of a vector signal generator with an arbitrary waveform generator, amplified by a typ. +45~dB microwave amplifier (Mini-circuit ZHL-16W-43S+), and passed through the stripline.
The AC external magnetic field is generated by the solenoid coil situated beneath the diamond, as presented in Fig.~\ref{fig:setup}(a).
A function generator and typ. +43~dB amplifier (Mini-circuit LZY-22+) to generate a sinusoidal wave voltage at 500.1~kHz are connected to the coil. The function generator is triggered by the same arbitrary waveform generator that synchronizes the phase at the start of the synchronized readout.
The clocks on all instruments are locked by a rubidium clock.

\subsection{Calibration}\label{subsec:pretest}
We check the linearity between the applied voltage amplitude of the function generator and the resulting field amplitude $b_\text{ac}$ before the main experiment.
We measure $b_\text{ac}$ with the AC magnetometry by sweeping the interpulse delay $\tau$ of the ordinary CP sequence~\cite{kotler2013nonlinear}.
We set the $\pi$ pulse duration to $t_{\pi}=19.8$~ns and the number of $\pi$ pulses to $N=16$ for the CP sequence.
Only in this measurement do we randomize the phase of the AC magnetic field and deviate the period of the pulse sequence from that of the AC field.
The obtained spectra for applied voltage amplitude $0.0, 2.0, 5.0, 7.0, 10.0, 15.0$, and $20.0~\si{\milli\volt}_\text{pp}$ are shown in Fig.~\ref{fig:setup}(b).
The horizontal axis is the detection frequency $1/2\tau$, and the vertical axis is the transition probability of the NV center.
As the voltage is increased, a signal appears near 500.1~kHz.
We fit the data with the following analytical model~\cite{degen2017quantum},
\begin{align}
P_0 
&= \frac{1}{2}\left(1-J_0(|W_a|\gamma b_{ac} N\tau)\right),\\
W_a &= \frac{\sin(\omega_\text{ac}N\tau/2)}{\omega_\text{ac}N\tau/2} \qty(1 - \frac{1}{\mathrm{cos}(\omega_\text{ac}\tau/2)} ),
\label{eq:s15}
\end{align}
where $J_0$ is the 0-th order Bessel function.
The fitted spectra are shown as the solid lines in Fig.~\ref{fig:setup}(b).
The model nicely explains the experimental data for all the input voltage amplitudes.
Note that the small difference between the experimental data and the fitting at around 300~kHz is due to the effect of $^{13} \mathrm{C}$ nuclear spins near the present NV center.
In order to obtain $b_\text{ac}$ accurately, we exclude this region from the fitting.
The obtained relationship between the applied voltage and the generated magnetic field amplitude is shown in Fig.~\ref{fig:setup}(c).
We get excellent linearity, yielding a conversion factor of $0.381\pm 0.004~\si{\micro\tesla/\milli\volt}_\text{pp}$.

\subsection{Protocol}\label{subsec:protocol}

\begin{figure}[hbtp]
\includegraphics[width=\hsize]{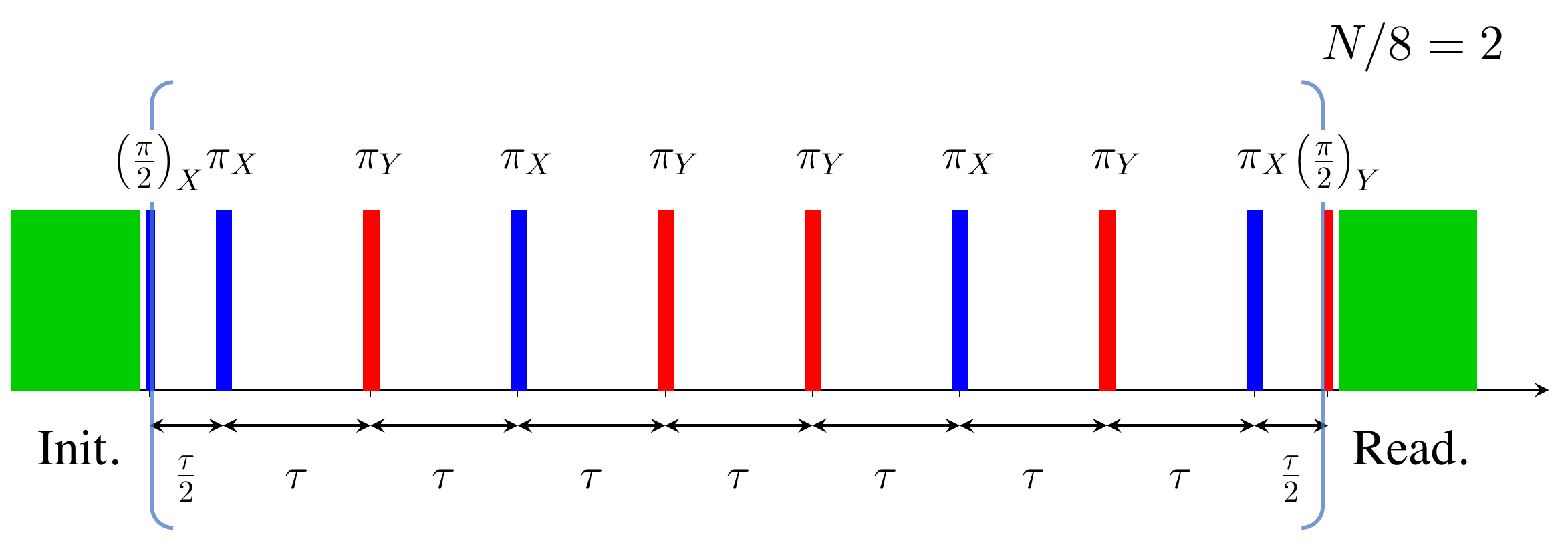}
\caption{XY8 pulse sequence used in our experiment ($N = 16$).}
\label{fig:xy8-diagram}
\end{figure}

Throughout our experiment, we use the same NV center and set the duration and the number of $\pi$ pulses to the same value as the ones we use in the prerequisite experiment
($t_{\pi}=19.8$~ns and $N=16$, respectively).
In this situation, we set the interpulse delay to $\tau = \pi/\omega_{\mathrm{ac}}$ to realize the Floquet state of $N/2=8$ cycles of the CP sequence.
We observe the response as formulated in Section ~\ref{subsec:Floquet_state_driven_by_CP_sequence}.
The AC magnetic field of $\omega_{\mathrm{ac}}=2\pi\times500.1$~kHz is generated by applying an AC voltage to the solenoid.
By increasing the voltage amplitude from $0.4\si{\milli\volt}_\text{pp}$ to $300\si{\milli\volt}_\text{pp}$, 
the AC magnetic field with an amplitude up to $\sim 120~\text{{\si{\micro \tesla}}}$ is estimated to be produced, which is far beyond the small-amplitude regime, $b_\text{ac} \approx \frac{\omega_\text{ac} }{2N \gamma} \times \frac{\pi}{2}= 877$~nT [Eq.~\eqref{baclin}].
We set the period of the synchronized readout as $\mathrm{{t_L=20.0~\si{\micro\second}} }$ and measure $335.5$ seconds with five-fold integration.
In this condition, $k$th harmonics should appear at a DFT frequency of $k\times100$~Hz (see Appendix A).
As mentioned in Section~\ref{sec:floquet}, the overall dynamics do not depend on the phase cycles of pulse operations.
We thus set the phase cycle as XY8 to reduce experimental pulse error. Figure~\ref{fig:xy8-diagram} shows the phase cycle we adopt in our experiment.

\section{Results}\label{sec:results}

\begin{figure}[tbp]
  \includegraphics[width=\hsize]{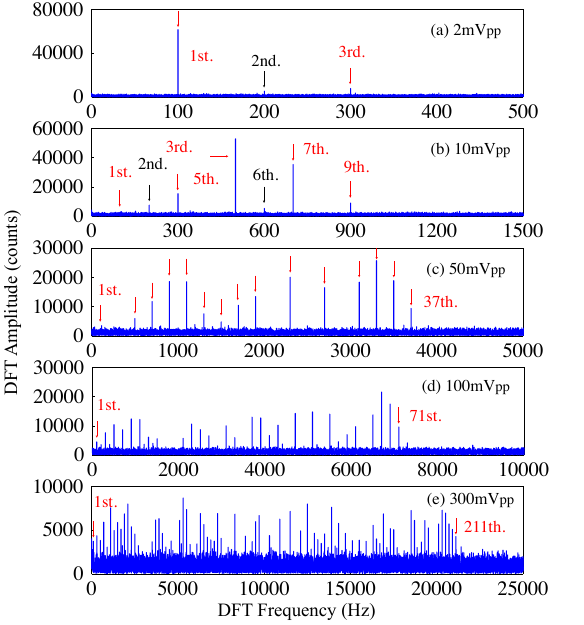}
  \caption{
  DFT spectra at several AC voltage amplitudes. 
  (a) $2~\si{\milli\volt}_\text{pp}$, (b) $10~\si{\milli\volt}_\text{pp}$, (c) $50~\si{\milli\volt}_\text{pp}$, (d) $100~\si{\milli\volt}_\text{pp}$, and (e) $300~\si{\milli\volt}_\text{pp}$.
  A finite vertical offset of about 3,000 counts due to photon shot noise is commonly present in all the data.
  }\label{fig:spectrum}
\end{figure}

\begin{figure*}[!tbp]
  \includegraphics[width=.95\hsize]{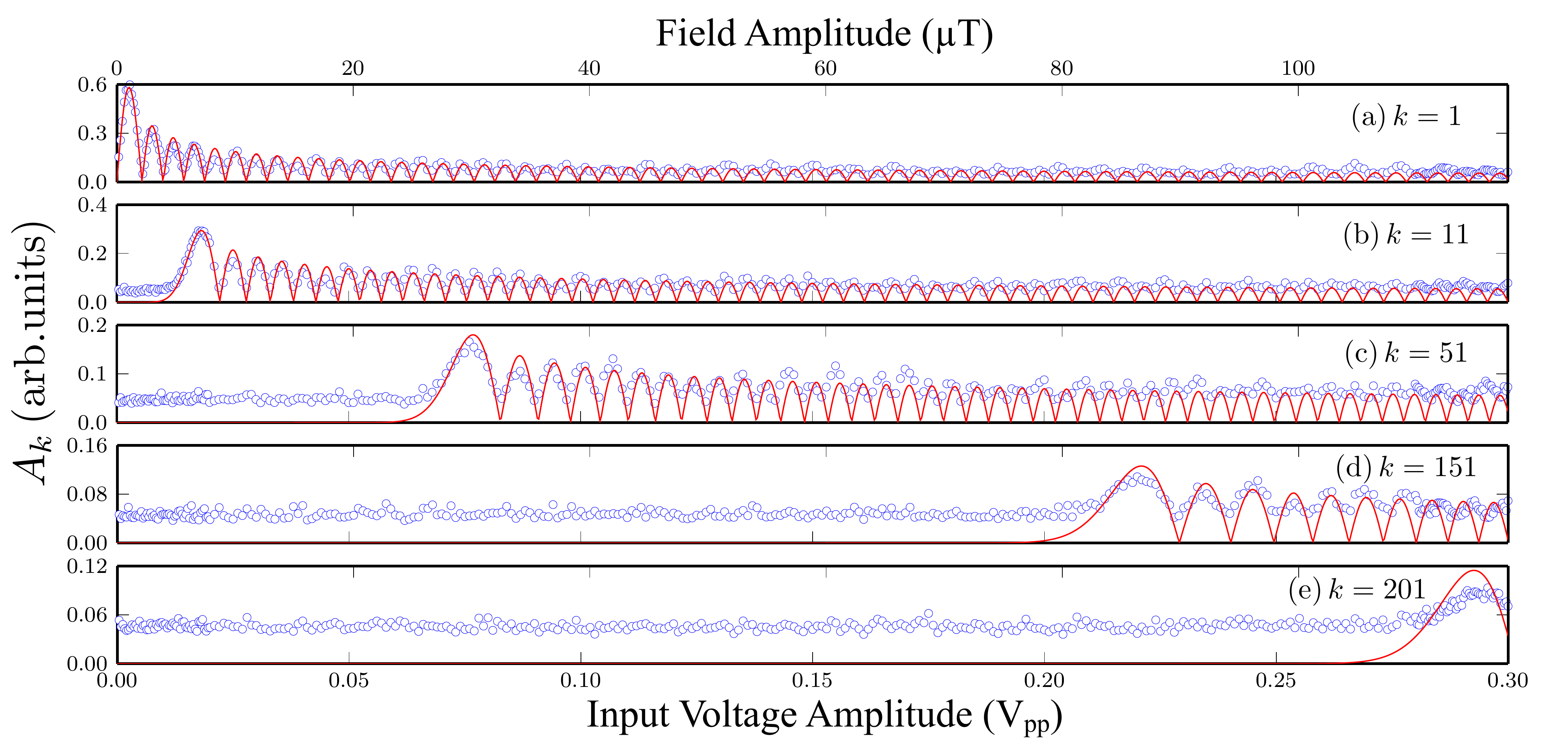}
  \caption{
  Input voltage amplitude dependence of the peak amplitude for (a) 1st, (b) 11th, (c) 51st, (d) 151st, and (e) 201st harmonics. The top axis show the magnetic field amplitude estimated from the fitting coefficient $0.392~{\si{\micro\tesla/\milli\volt}_\text{pp}}$ (see main text).
  }\label{fig:Ak-bessel}
\end{figure*}

\subsection{DFT spectra at each AC field amplitude}
Figure~\ref{fig:spectrum}(a) shows the DFT spectrum obtained when the voltage amplitude to generate AC magnetic field is set as small as $2~\si{\milli\volt}_\text{pp}$.
A large peak at 100~Hz and a small peak at 300~Hz correspond to the harmonics of $k=1$ and $k=3$, respectively.
Note that the signal is undersampled due to the time resolution of a finite measurement window (see Appendix ~\ref{sec:undersample}).
The absence of peaks higher than $k=3$ tells that the AC field amplitude is sufficiently small and the system is in the linear regime.
This is a condition similar to those reported in the previous studies using the synchronized readout~\cite{schmitt2017submillihertz, boss2017quantum, glenn2018high-resolution}.
Additionally, we see a peak at 200~Hz smaller than the peak at 300~Hz.
This even harmonic is caused by a pulse error (see Appendix C).
The present high-frequency resolution achieved by the  synchronized readout enables us to detect such a small signal.

\subsection{Subharmonic signals depending on field amplitude}
We then show the modulation voltage amplitude dependence of the DFT spectra.
Figures~\ref{fig:spectrum}(a), (b), (c), (d), and (e) show the DFT spectra when voltage amplitudes of $2~\si{\milli\volt}_\text{pp}$, $10~\si{\milli\volt}_\text{pp}$, $50~\si{\milli\volt}_\text{pp}$, $100~\si{\milli\volt}_\text{pp}$, and $300~\si{\milli\volt}_\text{pp}$ are applied, respectively.
As seen in these figures, odd harmonics appear up to higher orders successively as the amplitude increases.
In particular, even a harmonic peak for $k=211$ is observed in the spectrum for $300~\si{\milli\volt}_\text{pp}$ in Fig.~\ref{fig:spectrum}(e).
Such a higher-order response has not been investigated with this high precision before. A finite offset of about 3,000 counts due to photon shot noise is commonly present in all the data in Figs.~\ref{fig:spectrum} (a)--(e).

Now, we quantitatively examine the amplitude of each harmonic by comparing it to the Bessel function [Eq.~\eqref{eq:Ak}].
Figures~\ref{fig:Ak-bessel} (a), (b), (c), (d), and (e) show the results for 1st, 11th, 51st, 151st, and 201st harmonics, respectively.
The blue circles show the DFT amplitudes obtained by fitting the DFT spectra normalized to the state probabilities by the experimentally calibrated NV center's photoluminescence intensity (see Appendix B).
The constant offset of about 0.05 present in each figure is an artifact in the peak analysis due to the photon shot noise mentioned above.
The solid red lines show the analytical results fitted by Eq.~\eqref{eq:Ak}, reasonably assuming that the voltage amplitude and AC magnetic field amplitude are proportional and using only its proportionality coefficient as a fitting parameter.
The obtained coefficient is $0.392~{\si{\micro\tesla/\milli\volt}_\text{pp}}$, almost consistent with that obtained from the prerequisite test ($0.381\pm 0.004~\si{\micro\tesla/\milli\volt}_\text{pp}$) in Section~\ref{subsec:pretest}. Slight deviation ($\sim2.9\%$) is possibly due to the quasi-periodic distortion observed in experimental signal in Fig.~\ref{fig:Ak-bessel}, which will be discussed in the next subsection.

We find that the oscillation period and amplitude of both experimental and theoretical [Eq.~\eqref{fig:Ak-bessel}] oscillation are in general agreement, even up to higher-order harmonics exceeding 200.
These results mean that the Floquet state of the pulse-driven NV center is maintained even in large-amplitude modulation.
It is also significant that each DFT amplitude is consistent with theory without any artificial normalization; it indicates that the synchronized readout contributes to a highly quantitative measurement.

The AC field amplitude in the present experiment ranges from $157~\mathrm{nT}$ at minimum to $118~\text{{\si{\micro \tesla}}}$ at maximum, confirming that our investigation is systematic over a wide amplitude range, from a regime near the linear response to one that is highly nonlinear and nonequilibrium.
Although similar experiments in previous studies have observed the response to be a Bessel function~\cite{mizuno2020simultaneous} and the appearance of multiple harmonics~\cite{barson2021nanoscale}, we find that the peak amplitudes are quantitatively consistent over a significantly more extensive amplitude range than those.
While large-amplitude modulation in different physical phenomena can give dynamics that exhibit Bessel functions~\cite{saito2006parametric,weitenberg2021tailoring}, Bessel functions of as high as 200 orders have never been observed experimentally. The synchronized readout for the NV center is relevant in addressing Floquet state dynamics in such a large-amplitude modulation.

\begin{figure}[h]
\includegraphics[width = \linewidth]{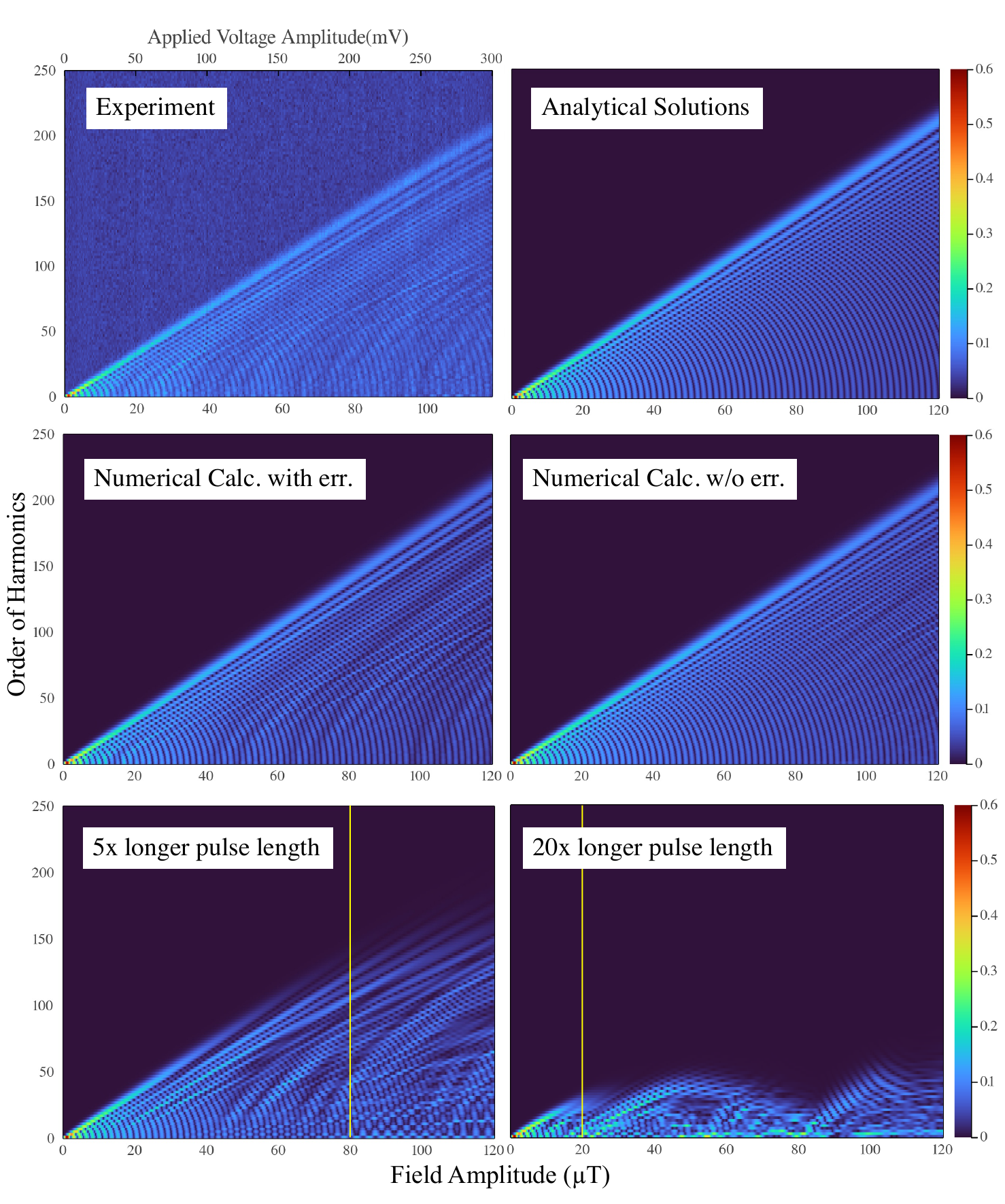}
\caption{
  Two-dimensional intensity plots of the peak amplitude as functions of the AC field amplitude and the harmonics $k$ up to 250.
  The color scale is common for all the panels (a)--(f).
  (a) Experimental result. 
  (b) Analytical result [Eq.~\eqref{eq:Ak}]. 
  (c) Numerical result including finite pulse duration with the duration error of 4~\% ($-0.8$~ns).
  (d) Numerical result including finite pulse duration without the duration error.
  (e) Numerical result  considering $5\times$ longer duration than the present experiment ($t_{\pi} = 100$~ns).
  (f) Numerical result  considering $20\times$ longer duration than the present experiment ($t_{\pi} = 400$~ns).
}\label{fig:results}
\end{figure}

\subsection{Comparison between experiments and theoretical calculation}\label{subsec:comparison}

We compare all of the experimentally observed harmonics with the theoretical calculations in terms of the AC field amplitude dependence.
Figure~\ref{fig:results}(a) shows the intensity plot of the experimental result as functions of the field amplitude and the order of harmonics $k$.
Figure~\ref{fig:results}(b) shows the corresponding $\abs{A_k}$, deduced from Eq.~\eqref{eq:Ak}.
These figures are almost consistent, indicating that Eq.~\eqref{eq:Ak} can, in principle, explain all harmonic behaviors observed in this experiment.

A more careful comparison reveals irregular fluctuations in the experimental intensity [Fig.~\ref{fig:results}(a)], which is not present in Fig.~\ref{fig:results}(b).
We compare this behavior with a more detailed physical model that includes finite pulse duration and errors.
Figures~\ref{fig:results}(c) and (d) show the numerical results considering finite pulse duration with and without the duration error of 4~\% ($-0.8$~ns), respectively.
Both show irregular fluctuations, and in particular, the result in Fig.~\ref{fig:results}(c) agrees very well with the experimental results [Fig.~\ref{fig:results}(a)].
These observations suggest a slight inevitable pulse error in the experiment.
This fact also explains the appearance of the small even-order harmonics in Fig.~\ref{fig:spectrum}.
Such a small pulse error can appear in the rise and fall times of the rectangular pulses used in the CP sequence.

Further numerical calculations show that the bandwidth of the pulse excitation limits the measurable amplitude range.
Figures~\ref{fig:results}(e) and (f) show the numerical calculations for $t_{\pi} = 100$~ns and $t_{\pi} = 400$~ns, respectively.
The respective pulse durations are 5 and 20 times longer than that used in the present experiment ($t_{\pi} = 19.8$~ns).
As shown in Figs.~\ref{fig:results}(e) and (f), as the pulse duration lengthens, more irregular fluctuations appear, and higher-order harmonics are no longer observable.
This is because the bandwidth of the pulse excitation ($1/t_{\pi}$) gets narrower than the range of the NV center's resonance frequency modulation.
We observe that the amplitude of the higher-order harmonics is almost zero in the area where the AC field amplitude is large, i.e., larger than that around 80~{\si{\micro \tesla}} in Fig.~\ref{fig:results}(e) and around 20~{\si{\micro \tesla}}  in Fig.~\ref{fig:results}(f). There, the driving force by AC field exceeds the microwave pulse driving and breaks down the pulse-driven Floquet state.

\section{Discussion}\label{sec:discussion}
We discuss two significant implications of the present result for Floquet engineering.
First, based on the above development, we can largely extend the modulation amplitude range available for Floquet engineering, such as quantum sensing~\cite{lang2015dynamical,stark2017narrow,saijo2018ac,morishita2019extension,meinel2021heterodyne}. The demonstrated method is advantageous in observing dressed states~\cite{belthangady2013dressed,morishita2019extension,stark2017narrow,stark2018clock} and many-body states formed by interactions with surrounding spins~ ~\cite{choi2017observation,randall2021many}.
As a system with robust Floquet states, the pulse-driven NV center can also be helpful in evaluating the effects of finite pulse duration and error in general two-level systems~\cite{loretz2015spurious, boss2016one-, ishikawa2019finite-pulse-width, lang2017enhanced, hengyun2020quantum}. 
Second, exploring the non-Hermitian Floquet dynamics is a promising direction.
In the present experiment, the CP sequence readout completely initializes the NV center, quenching the Floquet state.
Conversely, the synchronized readout with partial initialization or weak measurement of the NV center and surrounding nuclear spins~\cite{pfender2019high,cujia2019tracking} would serve as a platform to investigate non-Hermitian Floquet dynamics~\cite{kessler2021observation,beatrez2021floquet}.

\section{Conclusion}\label{sec:conclusion}
We precisely observed the dynamics of the NV center's Floquet state driven by the CP sequence in large-amplitude modulation as the higher-order harmonics up to 211 by using the synchronized readout technique.
We have thus established the relevance of the high precision of the synchronized readout in Floquet engineering. 
The nonlinear response of the Floquet state to the magnetic field is quantitatively reproduced by numerical simulations, including the effects of finite pulse duration and error.
This study further enhances the potential of the NV center as an ideal platform for Floquet engineering.

Finally, we consider potential applications of the present result to quantum sensing. 
Our experiment has confirmed that multiple peaks sharply change according to the Bessel functions in response to changes in the magnetic field amplitude, and this behavior is maintained even when a large amplitude magnetic field is applied. 
This fact means that the high magnetic field sensitivity of the synchronized readout~\cite{glenn2018high-resolution} is, in principle, sustained over a wide dynamic range. Thus we propose that this method is particularly suitable for physical property measurements using NV centers; We give two specific examples of its applications. 
One is the precise measurement of minute stray field. In recent years, atomically thin layered materials~\cite{zhang2021two-dimensional,xue2021recent,mak2019probing} have been studied extensively, and our scheme applies to the precise measurement of various sorts of magnetization, i.e., ferromagnetism, anti-ferromagnetism, and diamagnetism, of these materials. 
The other is its application to spin glass~\cite{binder1986spinglass,taniguchi2020spin}, a representative of systems where fluctuations play an essential role. Our technique is critical in understanding magnetic fluctuations near the spin-glass transition through quantitative measurements of the frequency spectrum of the complex susceptibility.

\section*{Acknowledgement}
We thank Takashi Oka and Naoto Tsuji for the helpful discussion and Takuya Isogawa for assistance in early experiments. 
This work was supported by Forefront Physics and Mathematics Program to Drive Transformation (FoPM), a World-leading Innovative Graduate Study (WINGS) Program, the University of Tokyo, and by Grants-in-Aid for Scientific Research Nos.~JP19H00656, JP19H05826, JP20K22325, JP22K03524, JP18H01502, JP22H01558, and JP19H02547, and by Q-LEAP (No.~JPMXS0118067395), and by Center for Spintronics Research Network, Keio University.

\section{Appendix}
\appendix

\section{Undersampling}\label{sec:undersample} 
We explain why the target AC field frequency ($500.1$~kHz) is different from the measured frequency of harmonics (multiples of $100$~Hz). We used a sequence whose total duration $t_\text{L}$ is 20~{\si{\micro \second}}, including initialization, pulse manipulation, and readout. In this scope, since the measurement interval is longer than the period of the target AC field, the frequency of the signal should stroboscopically get undersampled below the Nyquist frequency $\omega_\text{NY} = \pi/t_\text{L}$. In the experiment, the detected frequency $\omega_\text{ac}^\text{NY}$ should take the following value
\begin{equation}
    \omega_\text{ac}^\text{NY} = \omega_\text{ac} -  M \omega_\text{NY},
    \quad \text{where } M = \left\lfloor \frac{\omega_\text{ac}}{\omega_\text{NY}}\right\rfloor.
\end{equation}
The peaks of the general harmonic signal should therefore appear at multiples of $100~\text{Hz}~ [= 500.1~\text{kHz} - 20 \times 25~\text{kHz}~(= 1/2/20~\text{{\si{\micro \second}}}^{-1})]$.

\section{Processing of DFT signals} \label{sec:dft}
We explain the procedure to extract the amplitude of the DFT peaks. 
We perform the fitting of the DFT spectrum, taking it into account that the peaks of the data have a finite width.
We consider that the measurement time window is finite ($t \in [0,T_\text{tot}]$ , $T_\text{tot}$ is the total duration of synchronized readout.)
We assume the time series data as following simplified form:
\begin{align}
p(t) = A e^{i(\omega t + \phi)}, \, t \in [0,T_\text{tot}],\, A,\phi \in\mathbb{R}
\end{align}

This signal can be Fourier transformed as,

\small
\begin{align}
\mathcal{F}[p](\xi) &= \int_0^{T_\text{tot}} p(t) e^{-i\xi t}dt \notag\\
&= \frac{A T_\text{tot}}{4}\exp(i\phi -i(\xi  - \omega)T_\text{tot}/2) \mathrm{sinc}\qty[(\xi - \omega)T_\text{tot}/2],
\label{eq:sinc}
\end{align}
\normalsize
where $\mathrm{sinc}(x) \stackrel{\text{def}}{=} \sin x /x$.
Eq.~\eqref{eq:sinc} means that the DFT peak spreads over a finite width in the form of a sinc function.
We estimate the DFT peak amplitude by fitting the DFT spectrum with the peak shape represented in Eq.~\eqref{eq:sinc}.

\begin{figure}[htbp]
\includegraphics[width=\hsize]{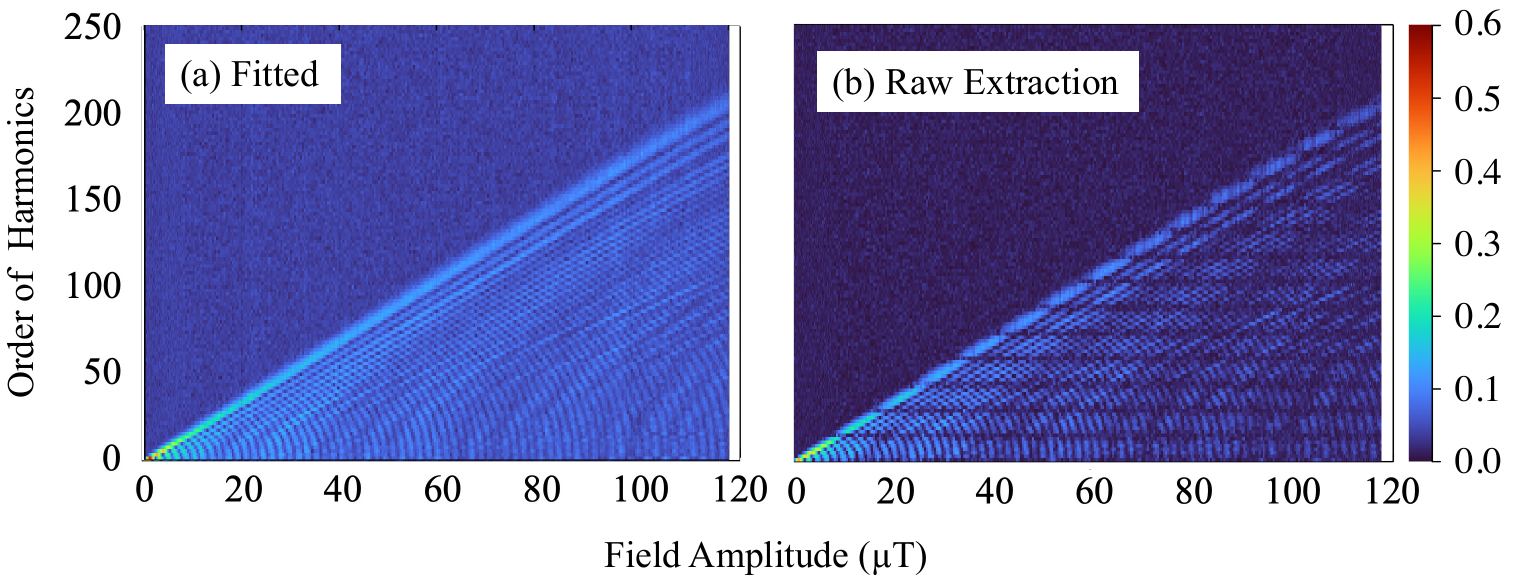}
\caption{\label{fig:schematic}
Intensity map of the DFT peak amplitude.
(a) Analyzed result with our fitting method, which is the same figure shown in Fig.~\ref{fig:results}(a). 
(b) The peak amplitude of the closest frequency point for $k~$th harmonic.
}
\end{figure}

Figure~\ref{fig:schematic}(a) shows the intensity plot obtained using the above method [the same as Fig.~\ref{fig:results}(a)]. On the other hand, Fig.~\ref{fig:schematic}(b) shows the plot obtained simply by extracting the data at the frequency index closest to the $k$~th peak frequency in the DFT spectrum.
In Fig.~\ref{fig:schematic}(a), a smooth image is obtained, but in Fig.~\ref{fig:schematic}(b), there is an artifact where the peak amplitude becomes smaller in a certain period over $k$.
This is an artifact due to the mismatch between the exact peak frequency and the frequency point of the DFT spectrum limited by the frequency resolution $1/T_{\mathrm{tot}}$.
The above-described method allows accurate determination of the peak amplitude.

After obtaining the DFT peak values, we normalize the time series data obtained from the synchronized readout for quantitative comparisons with theoretical models.
Using the photoluminescence intensity $I_0$ (count/sec) of the $m_S=0$ state and $I_1$ (count/sec) of the $m_S=-1$ state at the NV center as reference data, we normalize the DFT amplitudes (count) to the dimensionless value of $A_k$ ~\cite{misonou2020construction}.
Specifically, the conversion of the scaling is given as
\begin{align}
 \text{(DFT Amp.)} = A_k N_\text{seq}(I_0-I_1)t_\text{read}/2 .
\end{align}
$t_\text{read}$ is the readout duration (sec) and $N_\text{seq}$ is the total number of CP sequence repetitions.

\section{Effect of the pulse error and finite pulse duration}\label{sec:error}

We consider the effects of finite pulse duration.
First, we consider the pulse error in the $\pi/2$ pulse.
When the rotation angle of the pulse is $\pi/2 + \epsilon$, Eq.~\eqref{eq:p0} is rewritten as
\begin{equation}
P_0 \approx \frac{1}{2}(1 + \sin\phi_\text{acq} - \epsilon\cos\phi_\text{acq}).
\end{equation}
Compared to Eq.~\eqref{eq:p0}, the pulse error occurs as a cosine term.
This term appears as an effect that produces even-order harmonics of the DFT spectrum of the synchronized readout [see Fig.~\ref{fig:results}(a)].

Next, we consider the effects of finite pulse duration and error in $\pi$ pulses.
Since these effects can produce a variety of effects~\cite{loretz2015spurious, boss2016one-, ishikawa2019finite-pulse-width, lang2017enhanced, hengyun2020quantum}, we numerically investigate the situation to reproduce experimental observation.
Specifically, we solve the time-dependent Schr\"odinger equation Eq.~\eqref{eq:schr} of the total Hamiltonian $\hat{H}(t) = \hat{H}^{\mathrm{AC}} (t) + \hat{H}^{\mathrm{CP}} (t)$  [Eq.~\eqref{eq:htot}] with finite duration using the adaptive-step Runge-Kutta method.
All parameters including pulse phase cycle (XY8) of the Hamiltonian are based on experimental data except the true Rabi frequency. We show the results of these numerical simulations in Fig.~\ref{fig:results}.

\end{document}